\documentclass[fleqn,usenatbib]{mnras}
\usepackage{newtxtext,newtxmath}
\usepackage[T1]{fontenc}

\DeclareRobustCommand{\VAN}[3]{#2}
\let\VANthebibliography\thebibliography
\def\thebibliography{\DeclareRobustCommand{\VAN}[3]{##3}\VANthebibliography}

\usepackage{siunitx}
\usepackage{graphicx}	
\usepackage{amsmath}	

\usepackage{adjustbox}
\usepackage{enumitem}
\usepackage{tikz}
\usepackage{siunitx}
\usepackage{color,soul}

\usetikzlibrary{shapes,arrows,positioning,fit}

\title[A ML-based cool dwarf selection technique]{Atmospheric model-trained machine learning selection and classification of ultracool TY dwarfs}

\author[Ankit Biswas]{
Ankit Biswas,$^{1}$
\\
$^{1}$North Carolina School of Science and Mathematics, 1219 Broad St, Durham, NC 27705\\
}

\date{Accepted 2025 June 29. Received 2025 June 19; in original form 2024 March 11
}

\pubyear{2024}

\begin{document}
\label{firstpage}
\pagerange{\pageref{firstpage}--\pageref{lastpage}}
\maketitle

\begin{abstract}

\noindent The T and Y spectral classes represent the coolest and lowest-mass population of brown dwarfs, yet their census remains incomplete due to limited statistics. Existing detection frameworks are often constrained to identifying M, L, and early T dwarfs, owing to the sparse observational sample of ultracool dwarfs (UCDs) at later types. This paper presents a novel machine learning framework capable of detecting and classifying late-T and Y dwarfs, trained entirely on synthetic photometry from atmospheric models. Utilizing grids from the \emph{ATMO 2020} and \emph{Sonora Bobcat} models, I produce a training dataset over two orders of magnitude larger than any empirical set of $>$T6 UCDs. Polynomial color relations fitted to the model photometry are used to assign spectral types to these synthetic models, which in turn train an ensemble of classifiers to identify and classify the spectral type of late UCDs. The model is highly performant when validating on both synthetic and empirical datasets, verifying catalogs of known UCDs with object classification metrics $>99\%$ and an average spectral type precision within $0.35 \pm 0.37$ subtypes. Application of the model to a $1.5^\circ$ region in Pisces and the UKIDSS UDS field results in the discovery of one previously uncatalogued T8.2 candidate, demonstrating the ability of this model-trained approach in discovering faint, late-type UCDs from photometric catalogs.
\end{abstract}


\maketitle
\section{Introduction}
\label{sec:intro}

One of the major obstacles in understanding the nature of T and Y dwarfs is the detection of adequately-large samples of such objects. In order to obtain a complete and pure corpus of such objects, robust color, magnitude, or similar selection criteria are required, which are inferred from existing samples of late ultracool dwarfs (UCDs). The efficacy of detecting UCDs has grown significantly in the last three decades, with near-infrared surveys such as the \emph{Two Micron All-Sky Survey} \citep[2MASS;][]{skrutskie2006two}, \emph{Wide-Field Infrared Survey Explorer} \citep[WISE;][]{wright2010wide}, and the \emph{UKIRT Infrared Deep Sky Survey} \citep[UKIDSS;][]{lawrence2007ukirt} enabling the detection of hundreds of new objects. In particular, the JHK, W1, and W2 bands employed in these surveys reveal the presence of CH$_4$ and H$_2$O features in UCD atmospheres \citep{mainzer2003using}. The faint magnitudes probed by these surveys have helped define later spectral types and lower effective temperatures, down to the $\le 600 K$ Y dwarfs \citep{delorme2008cfbds, cushing2011discovery}. Despite the advent of deeper observations in recent years however, only 32 Y dwarfs and 895 T dwarfs are listed in the SIMBAD \citep{wenger2000simbad} database as of June 2025\footnote{Note that while SIMBAD attempts to provide a complete corpus of objects listed in the literature, it is still incomplete and inhomogenous in its coverage.}. In the corpus of 525 L/T/Y dwarfs within 20 pc of the Sun considered by \cite{kirkpatrick2021field}, the completeness limit falls steeply at spectral types later then T8, down to 13 pc for Y0-1.5 dwarfs. At the latest spectral types, it becomes difficult to robustly determine the luminosity function of these objects due to limited counts. 

Since the current empirical catalog of late TY dwarfs is limited, they cannot act as a robust training set for detection and classification methods. For over two decades, 1D radiative-convective models have attempted to replicate the colors of TY dwarfs. These models simulate the thermal structure and spectral characteristics of brown dwarf atmospheres by solving the radiative transfer equations while accounting for convective energy transport. Typically, this is used to compute a set of models over a discrete grid of parameters, resulting in a model grid. Recently, two independent teams have developed state-of-the-art compatible atmospheric model grids, \emph{ATMO 2020} \citep{phillips2020new} and \emph{Sonora Bobcat} \citep{marley2021sonora}, which both incorporate non-equilibrium (NEQ) chemistry. The former presents an updated equation of state and improved treatments of alkali lines in model atmospheres, while the latter delves into cooler effective temperatures and explores a large range of metallicities with updated line opacities. Retrieval analysis, however, has recently shown that while atmospheric models effectively predict the atmospheric parameters of early-and-late T dwarfs \citep{line2015uniform}, there are systemic biases and distance-dependent degeneracies when doing the same for early Y dwarfs \citep{zalesky2019uniform}. \cite{marley2021sonora} similarly acknowledges consistent discrepancies between model fits and empirical observations for Y dwarfs, and \cite{phillips2020new} discusses the unexplained flux at $\lambda \sim \qty{4}{\micro\meter}$ for objects cooler than $700 \text{K}$ which like arises from processes not addressed by standard radiative-convective models. This paper investigates whether these models can be utilized as training data for classification of brown dwarfs despite these deviations. 

One of the most useful tools in recent years for the detection and classification of brown dwarfs has been machine learning (ML). With the advent of large all-sky catalogs of millions of objects, ML has proven to be very effective in parsing the multidimensional color-magnitude parameter space of astronomical objects. There have been a few cases of using ML in relation to brown dwarfs for either discovery or the determination of properties. \cite{feeser2022using} used ML to determine the physical properties of brown dwarfs using low-resolution spectra. On the photometric side, \cite{sithajan2023applied} tested random forests (RF), k-Nearest-Neighbors classifiers (k-NN), and a multi-layer perceptron (MLP) on their ability to distinguish M-dwarf subtypes. 

While these studies focus on analyzing identified brown dwarfs, this paper is focused on both the discovery of brown dwarfs and subsequent spectral type assignment. In a similar vein, \cite{bhavana2019classifier} identified L and T dwarfs in photometric catalogs with high completeness using an ensemble machine learning classifer composed of an artificial neural network (ANN) and two variations of a k-Nearest-Neighbors classifier (k-NN). More recently, \cite{gong2022applying} and \cite{gutierrez2022applying} have trained 2-part random forest classifiers on dwarf colors from \cite{skrzypek2016photometric} and \cite{best2018photometry} to separate L and T-type UCDs from background sources and classify them by spectral type. 

The challenge with using these methods to search for late T and early Y dwarfs is that they rely heavily on observational data for training, restricting machine-learning-based searches to well-cataloged spectral types. The aim of this paper is to identify TY dwarfs and assign them spectral types with the use of ensemble machine learning techniques trained on atmospheric models. Section ~\ref{sec:data} discusses the training data, which encompasses the various brown dwarfs and contaminant models. Section ~\ref{sec:ml} discusses the ensemble classifier this work develops for the discovery and classification of UCDs, and finally, Section ~\ref{sec:results} covers the accuracy of said classifier and the discovery of UCD candidates. All magnitudes are given in the Vega system.

\section{Data}
\label{sec:data}

This section describes the \emph{ATMO 2020} and \emph{Sonora Bobcat} 1D radiative-convective substellar atmosphere model grids and details their use in this work. Below are brief descriptions on each of the pertinent data-sets, their important characteristics, and how they were processed to create training data for the final ensemble classifier. 

\subsection{ATMO 2020 models}
\label{sec:atmo2020}

The \emph{ATMO 2020} models \citep{phillips2020new}, based on the group's earlier \emph{ATMO} code \citep{amundsen2014accuracy}, are cloudless atmospheres that present updated molecular line lists compared to previous models and improved modeling of alkali lines. The atmospheres are constrained within temperatures $200 \hspace{1mm} \mathrm{K} \le \mathrm{T_{eff}} \le 3000 \hspace{1mm} \mathrm{K}$, gravities $2.5 \hspace{1mm} \qty{}{\centi\meter\per\second\squared} \le \log{g} \le 5.5 \hspace{1mm} \qty{}{\centi\meter\per\second\squared}$, solar metallicity ($[M/H] = 0$), and eddy diffusion coefficients $-0.5 \hspace{1mm} \qty{}{\centi\meter\squared\per\second} \le \log(K_{zz}) \le 0.5 \hspace{1mm} \qty{}{\centi\meter\squared\per\second}$. \emph{ATMO 2020} presents three sets of model isochrones over the given parameter space; each possess a constant but different $\log(K_{zz})$ over the aforementioned range with steps of 0.5 dex.

\subsection{Sonora Bobcat models}
\label{sec:sonora}

The \emph{Sonora Bobcat} model set \citep{marley2021sonora} provides a large set of cloudless TY dwarf models that are compatible with the \emph{ATMO 2020} atmospheres. They are constrained within $200 \mathrm{K} \le \mathrm{T_{eff}} \le 2400 \mathrm{K}$ and $2.5 \hspace{1mm} \qty{}{\centi\meter\per\second\squared} \le \log{g} \le 5.5 \hspace{1mm} \qty{}{\centi\meter\per\second\squared}$. \emph{Sonora Bobcat} provides three branches of models, this time for -0.5, 0.0, and 0.5 metallicity (whereas the \emph{ATMO 2020} models are all at solar metallicity). 

To constrain the classifier to solely late-T and Y dwarfs, M, L, and early T dwarfs are filtered from both model data sets by requiring $\mathrm{T_{eff}} \le 1000$ K. This cut is supported by both theoretical and observational results \citep[][e.g.]{burrows2002theoretical, burgasser2002spectra, burrows2003beyond, burningham2008exploring, del2009physical}, which show that the early-late T dwarf transition lies at approximately 1000 K. 

\subsubsection{Interpolation}

\begin{figure}
  \resizebox{\hsize}{!}{\includegraphics{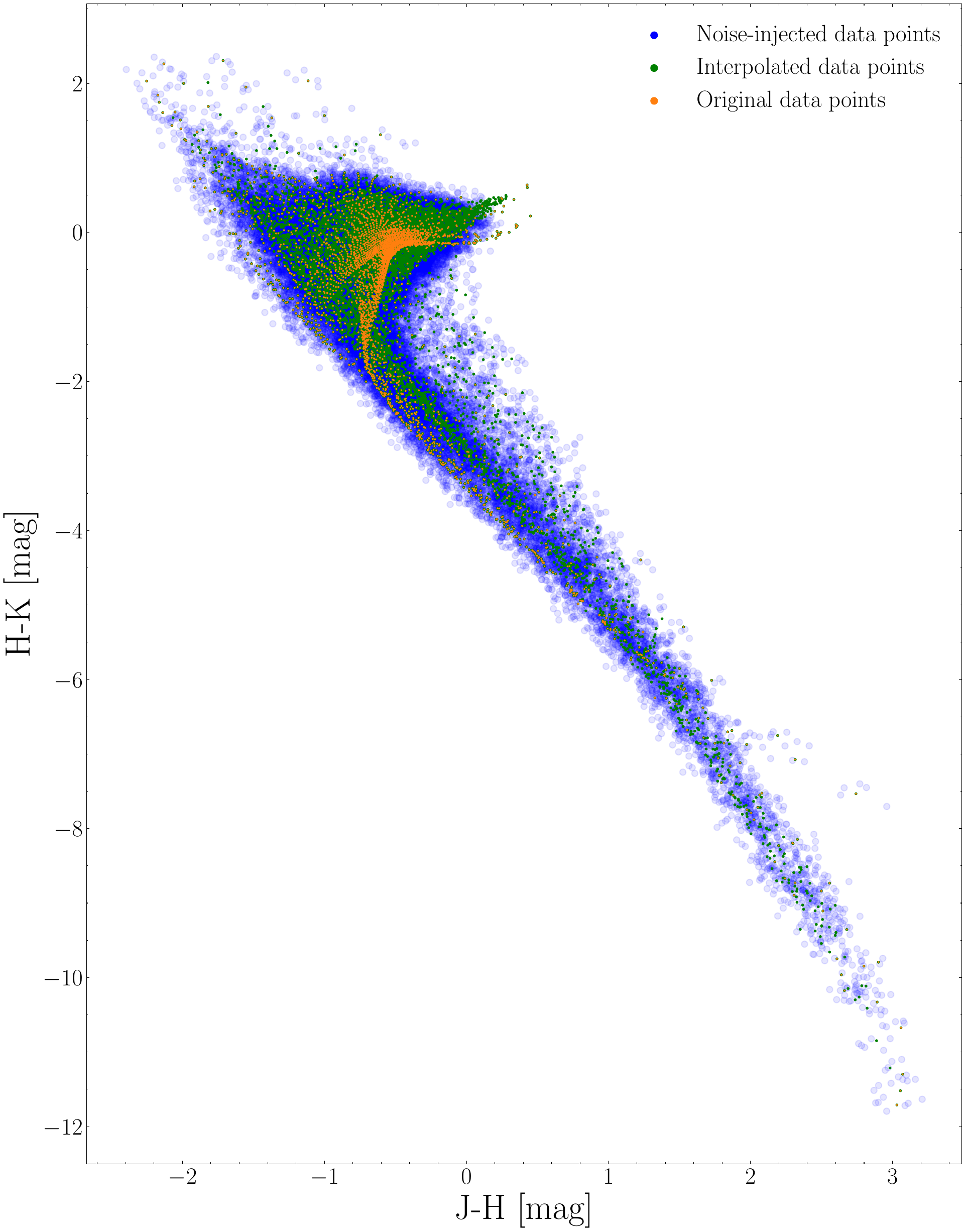}}
  \caption{Color-color diagram of the augmented total brown dwarf model set. Orange points indicate the original models, green points indicated interpolated models, and blue dots indicate the total noise-injected model set.}
  \label{fig:noiseinjected}
\end{figure}

In comparison to the $\sim 9000$ models of the filtered \emph{ATMO 2020} grid, the filtered \emph{Sonora Bobcat} grid contains $\sim 700$ models. To achieve adequate representation of non-solar metallicities in the combined model dataset, the Sonora grid must be augmented to reach a model count approximately equivalent to the \emph{ATMO 2020} dataset. Taking the $\mathrm{T_{eff}}$, $\log{g}$, mass, and [M/H] as the interpolation indexes, 8000 synthetic models are randomly generated in the parameter space with their YJHKW1W2W3 magnitudes interpolated linearly. To avoid undersampling in areas of the parameter space with low grid coverage, a random uniform sampling method is used to generate new points across the input domain (bounded between the minimum and maximum for each index parameter), rather than placing these new models at fixed intervals. The effect is shown in Figure ~\ref{fig:noiseinjected}, where the green interpolated points offer greater coverage than the original model grid where H-K $\ge 0$.

\subsection{Photometric classification}
\label{sec:photclass}

Training a classifier for spectral type assignment requires the models have labeled spectral types. This section details this process of fitting the synthetic model colors to spectral type polynomial relations from \citet{kirkpatrick2012further, kirkpatrick2019preliminary} and \citet{skrzypek2015photometric}. Hereafter, the filtered and interpolated \emph{ATMO 2020} and \emph{Sonora Bobcat} model sets are concatenated into one total dataset and will collectively be referred to as the "synthetic dwarfs" or any equivalent phrase. 

\subsubsection{Polynomial-fitting framework}
\label{sec:polynomials}

\emph{Photo-type}, developed by \citet{skrzypek2015photometric}, is a photometric classification procedure that employs polynomial-defined template colors to assign spectral types to observations based on a least-$\chi^2$ fit. This scheme is the base framework for the spectral type labeling of the synthetic dwarfs.

This work uses a modified version of the $\chi^2$ metric in \citet{skrzypek2015photometric}, 
\begin{equation}
    \lambda^2 = \sum_{b = 1}^{N_{b}} (m_{b} - m_{B, t} - c_{b, t})^2, 
\end{equation}
where $m_{B, t}$  is the inverse-variance weighted estimate of the magnitude in a reference band $B$, $m_{b}$ is the observed magnitude in a band $b$, and $c_{b, t}$ is the given template color between bands $b$ and $B$. Since the spectral type assignment is conducted with models with no associated photometric uncertainty, the colors are not divided by the standard deviation when fitting—thus, this work uses $\lambda^2$ instead of $\chi^2$ to avoid confusion. Similarly, the absence of model uncertainties or photometric errors makes the choice of the reference band $B$ largely arbitrary.  Ultimately, the fitting in this work utilizes H as the reference band due to its presence in both sets of relations fit to the models. 

\begin{figure*}
\centering
  \resizebox{\hsize}{!}{\includegraphics{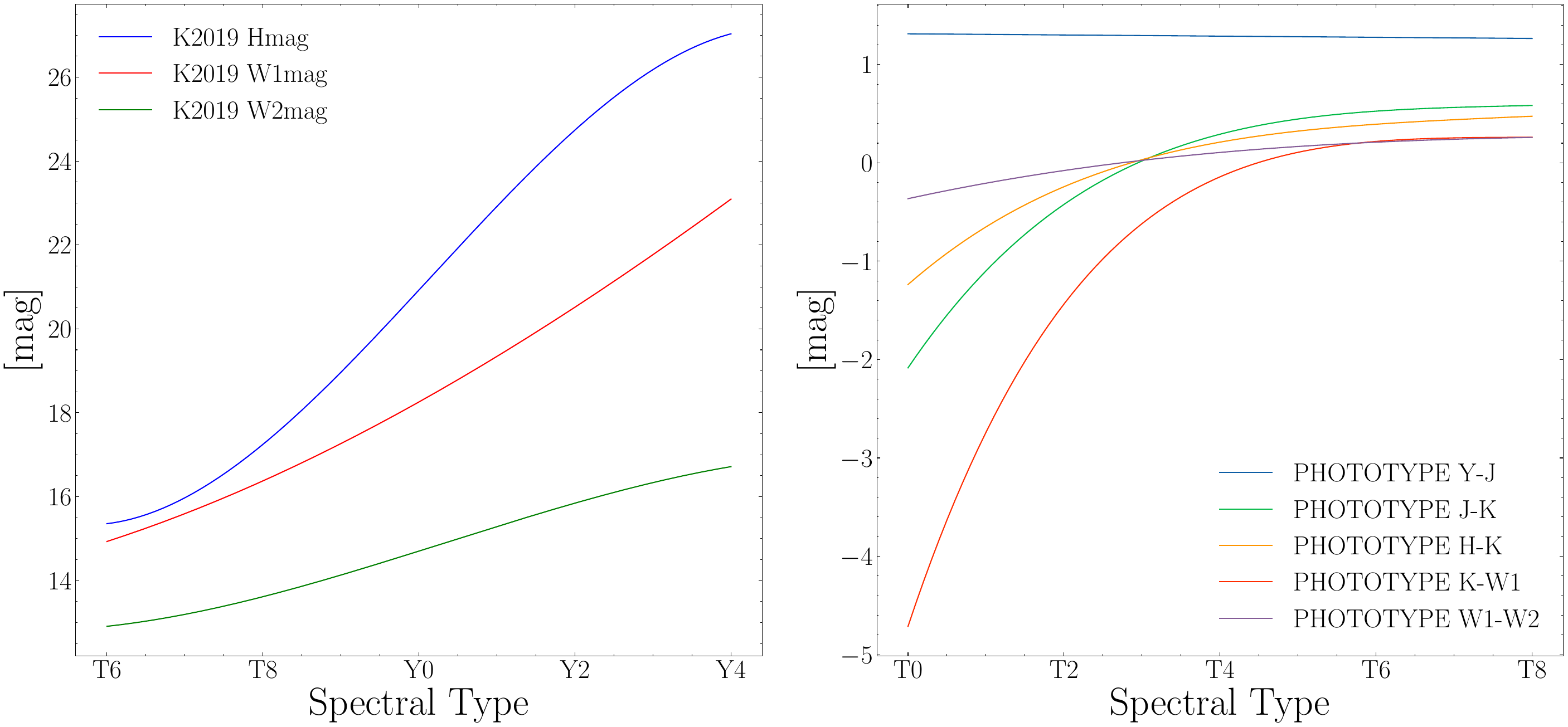}}
  \caption{Polynomials as described in Table ~\ref{table:2}. The ranges over which the polynomials are utilized are used as bounds for the polynomials in the plot. Left: Magnitude-based polynomials; Right: Color-based polynomials}
  \label{fig:polynomials}
\end{figure*}

In total, two "branches" of polynomial relations are used for $\lambda^2$ spectral type fitting: K19 and PHOTO, taken from \cite{kirkpatrick2019preliminary} and \cite{skrzypek2015photometric} respectively. These polynomials are empirically-derived fits from their respective censuses, ensuring observationally-supported spectral type classification. The polynomials from both branches are described in Table ~\ref{table:2} and shown in Fig. ~\ref{fig:polynomials}. Note that when fitting, the colors derived from these polynomial sets are subtracted such that all template colors are with respect to the reference band H. 

\begin{table*}
\caption{Polynomials of the form $y = \sum_{i = 0}^{n} c_{i}x^i$. SpT refers to spectral type, and $m_X$ refers to the magnitude or color in/between band(s) $X$. The K2019 and PHOTO polynomials are sourced from \protect\cite{kirkpatrick2019preliminary} and \protect\cite{skrzypek2015photometric} respectively.}
\label{table:2}
\centering
\begin{adjustbox}{width=\linewidth,center}
\begin{tabular}{ccccccccccc}
\hline
Branch & $y$ & $x$ & $c_0$ & $c_1$ & $c_2$ & $c_3$ & $c_4$ & $c_5$ & $c_6$ & Range \\
\hline
K2019 & $m_{H}$  & SpT & $37.0 \pm 4.5$ & $-8.67 \pm 1.56$ & $1.051 \pm 0.174$ & $-0.0345 \pm 0.0063$ & - & - & - & $T6 \le \mathrm{SpT} \le Y4$ \\
K2019 & $m_{W1}$ & SpT & $13.82 \pm 4.00$ & $-0.276 \pm 1.39$ & $0.0845 \pm 0.155$ & $-0.00124 \pm 0.00055$ & - & - & - & $T6 \le \mathrm{SpT} \le Y4$ \\
K2019 & $m_{W2}$ & SpT & $16.36 \pm 1.96$ & $-1.60 \pm 0.68$ & $0.211 \pm 0.075$ & $-0.00682 \pm 0.00272$ & - & - & - & $T6 \le \mathrm{SpT} \le Y4$ \\
K2019 & $m_{W3}$ & SpT & $10.7 \pm 8.9$ & $0.0320 \pm 0.2015$ & $0.0159 \pm 0.0115$ & $ - $ & - & - & - & $T6 \le \mathrm{SpT} \le Y4$ \\
PHOTO & $m_{Y-J}$ & SpT + 20 & 1.312 & -5.92e-3 & - & - & - & - & - & $T0 \le \mathrm{SpT} \le T8$ \\
PHOTO & $m_{J-H}$ & SpT + 20 & -2.084 & 1.17016 & -0.199519 & 0.01610708 & -5.93708e-4 & 7.94462e-06 & - & $T0 \le \mathrm{SpT} \le T8$ \\
PHOTO & $m_{H-K}$ & SpT + 20 & -1.237 & 0.69217 & -0.114951 & 9.46462e-3 & -3.61246e-4 & 5.00657e-06 & - & $T0 \le \mathrm{SpT} \le T8$ \\
PHOTO & $m_{K-W1}$ & SpT + 20 & -4.712 & 2.37847 & -0.444094 & 0.04074163 & -1.910084e-3 & 4.36754e-05 & -3.835e-07 & $T0 \le \mathrm{SpT} \le T8$ \\
PHOTO & $m_{W1-W2}$ & SpT + 20 & -0.364 & 0.17264 & -0.015729 & 4.8514e-4 & - & - & - & $T0 \le \mathrm{SpT} \le T8$ \\
\hline
\end{tabular}
\end{adjustbox}
\centering
\end{table*}

For each model atmosphere, two discrete spectral types are determined based off of the best $\lambda^2$ fits to the polynomial branches. Due to the varying number of polynomials per branch, the $\lambda^2$ metric is normalized to a reduced statistic $\lambda_r^2 = \frac{\lambda^2}{\nu}$, where $\lambda^2$ is normalized by dividing it by the degrees of freedom $\nu$ of the polynomial branch. This is analogous to the reduced chi-squared statistic. Following this approach, spectral types are assigned to each model by lowest-$\lambda_r^2$-fit to the polynomial-generated relations (see Figure ~\ref{fig:polynomials}. 

\begin{figure*}
  \resizebox{\hsize}{!}{\includegraphics{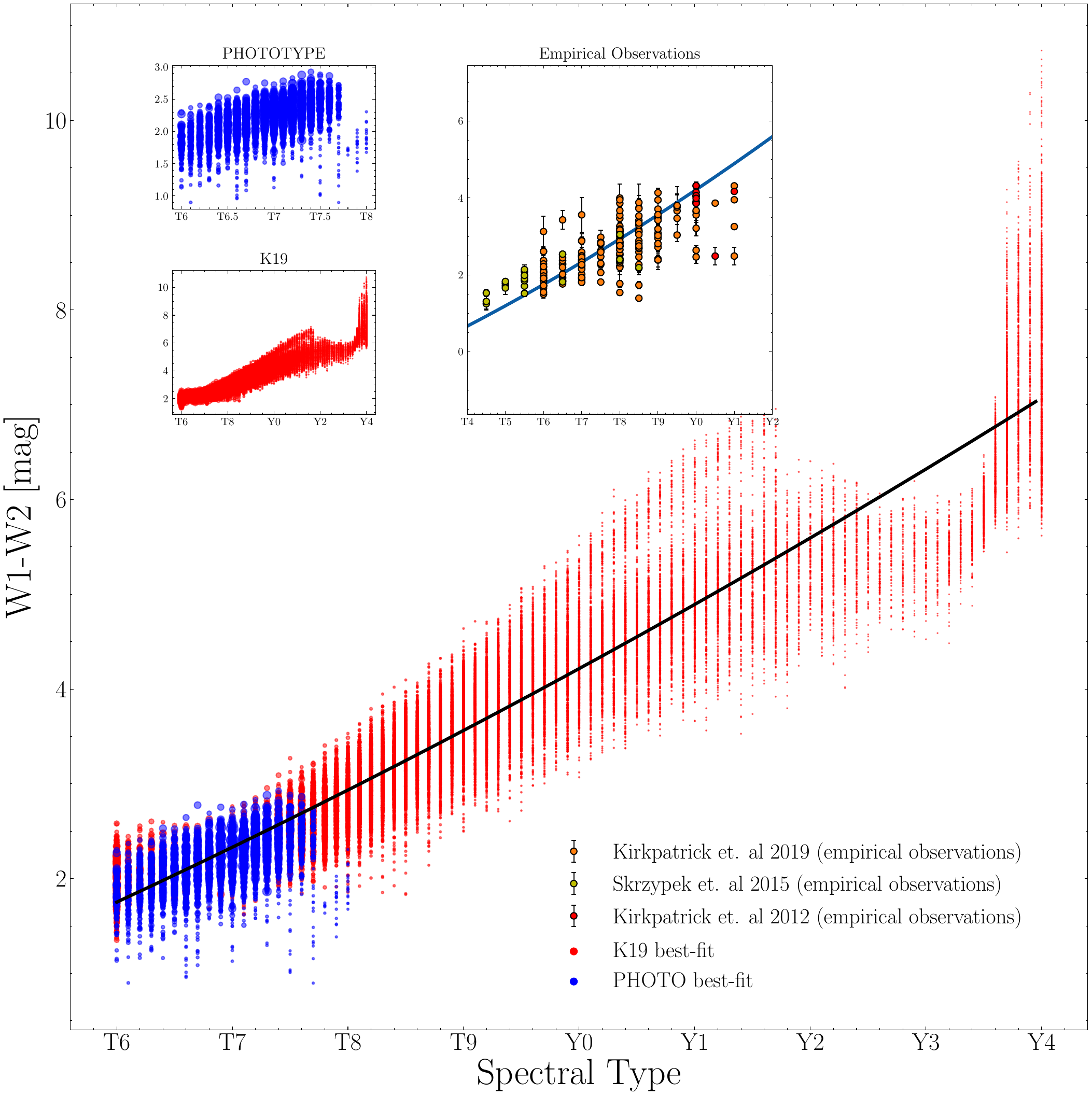}}
  \caption{Color-Spectral Type diagram of \emph{ATMO 2020} data set. Dark blue dots represent the PHOTOTYPE sequence and the red dots represent the K19 (see Table ~\ref{table:2}). Orange, yellow, and red points in the inset plot on the top right are empirical classifications from \citet{kirkpatrick2019preliminary, kirkpatrick2012further, skrzypek2015photometric}. Inset plots on the left contain snapshots of the same color-parameter spaces isolated with each of the indicated polynomial schemes. The equation of the line plotted in both the main figure and the inset plot, derived from a second-order fit to the model dwarfs, is $\text{W1-W2}=0.02332*\text{(SpT)}^2 + 0.1552*\text{(SpT)} + 0.2115$.}
  \label{fig:modelscatterplot}
\end{figure*}

A color-spectral type diagram of the combined model set is shown in Figure ~\ref{fig:modelscatterplot}. As shown in the inset plot with empirical dwarfs from \cite{kirkpatrick2012further, kirkpatrick2019preliminary, skrzypek2014brown}, the W1-W2 color trend of the models generally lines up with observations. However, the diagram also displays a large decrease in the number of models with spectral types classified between Y2-Y3.5 with a sharp uptick at Y4. This is more clearly visible in Figure ~\ref{fig:datahist}. This is likely due to uncertainties in the K19 polynomial scheme, which beyond Y2 is only bound by two dwarfs with upper-limit spectral type classifications \citep{kirkpatrick2019preliminary} and has the best-fit to all model photometry beyond $\sim$T8. As this work does not account for the uncertainty in these relations to avoid different treatment between the K19 and PHOTO branches–any classification variability is also expected to be accounted for in the noise injection procedure (see ~\ref{sec:dataaug})–the poor spectral type relation constraints $\ge$Y2 is the likely culprit of this artifact. Meanwhile, the spike at Y4 is a boundary artifact caused by a large number of $\ge$Y4 models in the model set. For this reason, this work considers any Y4 classification to be instead $\ge$Y4.

\subsection{Data augmentation}
\label{sec:dataaug}

The total model dataset ($\sim9600$ models) provides comprehensive coverage over a large grid of atmospheric and physical properties. However, the discrete nature of the models may also result in overfitting to well-defined intervals of values. As such, the model set is augmented via noise injection to mimic empirical photometric scatter. To achieve this, the combined model dataset is duplicated 14 times (arbitrarily chosen) and each passband magnitude value for each synthetic model is offset by a randomly-sampled value from a Gaussian distribution with standard error equal to the uncertainty for that band in \cite{skrzypek2016photometric}. The noise-injected dataset is displayed in Figure ~\ref{fig:noiseinjected}.

\subsection{Contamination modeling}
\label{sec:contaminants}

When searching for UCDs in a photometric catalog, the first step naturally is to discriminate between contaminants and actual target objects. Traditionally, the ML approach has been to use a 2-part classification system, the first classifier focused on discerning the target objects from background sources, and the second classifier focused on identifying the relevant property of the target object \citep{bhavana2019classifier, gong2022applying, gutierrez2022applying}. This work adopts this two-part approach. To train this first classifier, which will be referred to as an Object-type (O-type) classifier, I use a corpus of contaminant objects and their associated photometry from empirical/model catalogs. For each of the following catalogs, any missing photometric bands are added by cross-matching to the appropriate UKIDSS survey \citep{lawrence2007ukirt} and/or AllWISE \citep{wright2010wide,mainzer2011preliminary} using the \emph{Centre de Données astronomiques de Strasbourg} (CDS) cross-match service \citep{boch2012cds, pineau2020cds} with a radius of 3.5 arcseconds (approximately half the angular resolution of the W3 band, the lowest-resolution band used for classification).

The contaminant catalogs used are—Active Galactic Nuclei \citep[AGNs;][]{maddox2008luminous}, FGK stars \citep{sarmento2020derivation}, late M and L dwarf models above the 1000 K cut in Section~\ref{sec:sonora}, M dwarfs \citep{cook2017method}, M/L/early-T dwarfs from the UltracoolSheet compilation \citep{best2020volume, best2024volume}, NSL1 galaxies \citep{chen2017infrared}, red giant stars \citep{gontcharov2008red}, variable stars \citep{ferreira2015wfcam}, and Young Stellar Objects \citep[YSOs;][]{rebull2010taurus}. A comprehensive summary of these catalogs is shown in Table~\ref{table:contaminants}.

\begin{table}
\caption{Contaminant catalogs. AllWISE and UKIDSS columns indicate if cross-matching was required for those datasets.}              
\label{table:contaminants}      
\centering
\begin{adjustbox}{width=\linewidth,center} 
\begin{tabular}{lcccc}          
\hline
Type & Count & AllWISE & UKIDSS & Reference\\
\hline
AGNs & 367 & yes & no & \cite{maddox2008luminous}\\
FGK stars & 449 & no & yes & \cite{sarmento2020derivation}\\
Late M + L models & 1679 & - & - & \cite{phillips2020new,marley2021sonora}\\
M dwarfs & 7953 & yes & no & \cite{cook2017method}\\
M, L, and early T dwarfs & 1314 & no & no & \cite{best2020volume, best2024volume}\\
NSL1 galaxies & 319 & yes & no & \cite{chen2017infrared}\\
Red giants & 10624 & yes & yes & \cite{gontcharov2008red}\\
Variable stars & 221 & yes & no & \cite{ferreira2015wfcam}\\
YSOs & 86 & yes & yes & \cite{rebull2010taurus}\\
Background sources & 25935 & yes & no & - \\
\hline
\end{tabular}
\end{adjustbox}
\centering
\end{table}

Finally, to account for additional areas of the color space not covered by the specialized contaminant catalogs, one additional contaminant catalog was sampled from a $2^{\circ} \times 2^{\circ}$ square region centered at 11 32 50 2 +01 17 09.3. This cutout was chosen due to its full coverage in both UKIDSS and WISE data along with an absence of any detected UCDs. Combined with the specialized contaminant catalogs, it is hoped this random selection of background sources helps cover the entirety of the relevant color-magnitude space. 

\subsubsection{Class balancing}

As shown in Table ~\ref{table:contaminants}, there is significant class imbalance in the initial contaminant catalogs, where particular object types are overrepresented. To ensure that all classes are represented equally for training the classifier, each individual contaminant catalog is oversampled/undersampled into a new overarching contaminant set. This class balancing is conducted as follows: 
\begin{enumerate}
    \item Choose an arbitrary target number of objects/models per contaminant catalog. In this work, a target of 1000 objects/models is selected.
    \item If a catalog contains more objects/models than the target, randomly select the target number of objects/models from the catalog.
    \item If a catalog contains less objects/models than the target, duplicate the catalog until it equals or exceeds the target. 
\end{enumerate}

\begin{figure*}
\begin{tabular}{cc}

\includegraphics[width=\columnwidth]{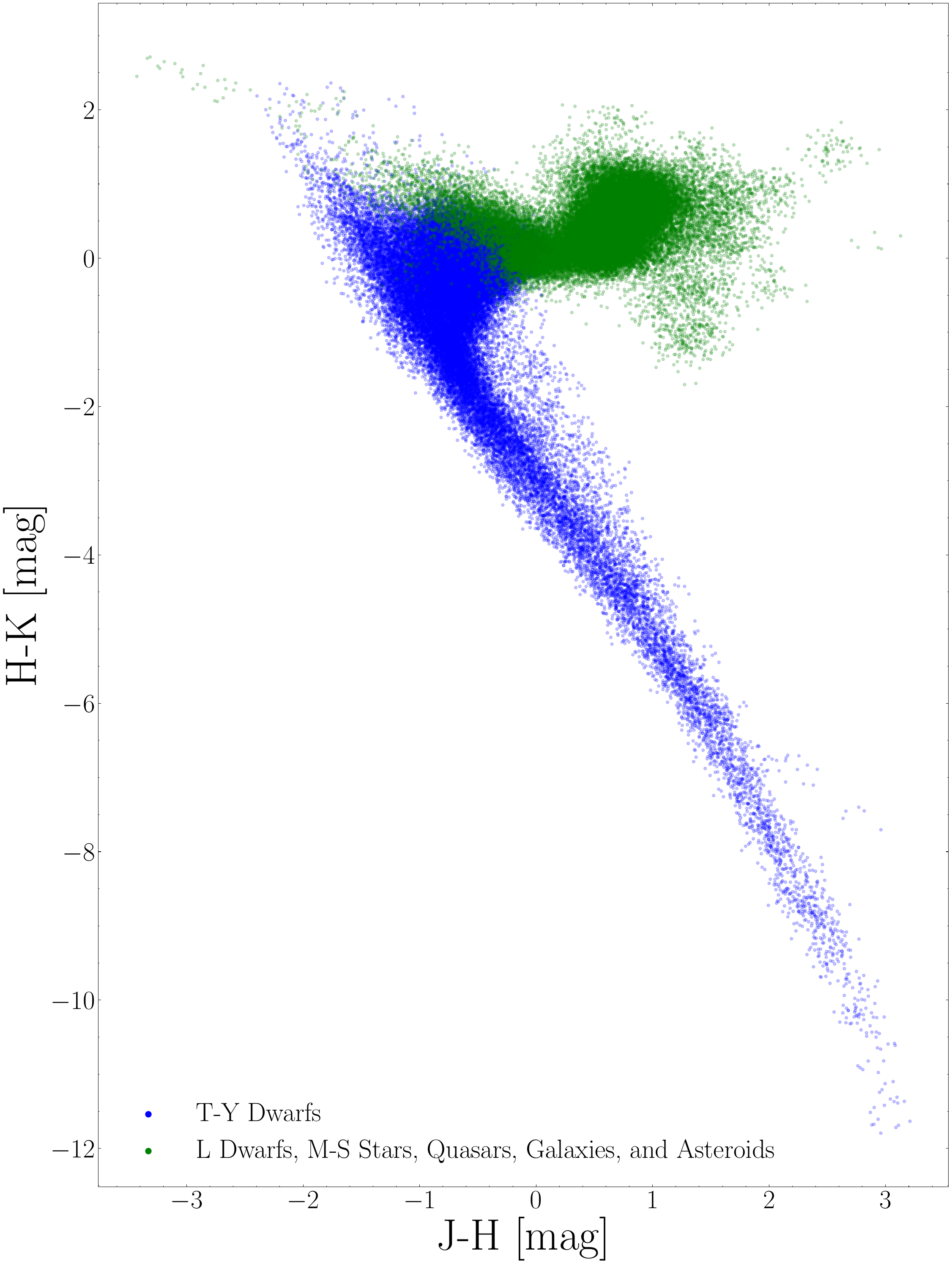}
&
\includegraphics[width=\columnwidth]{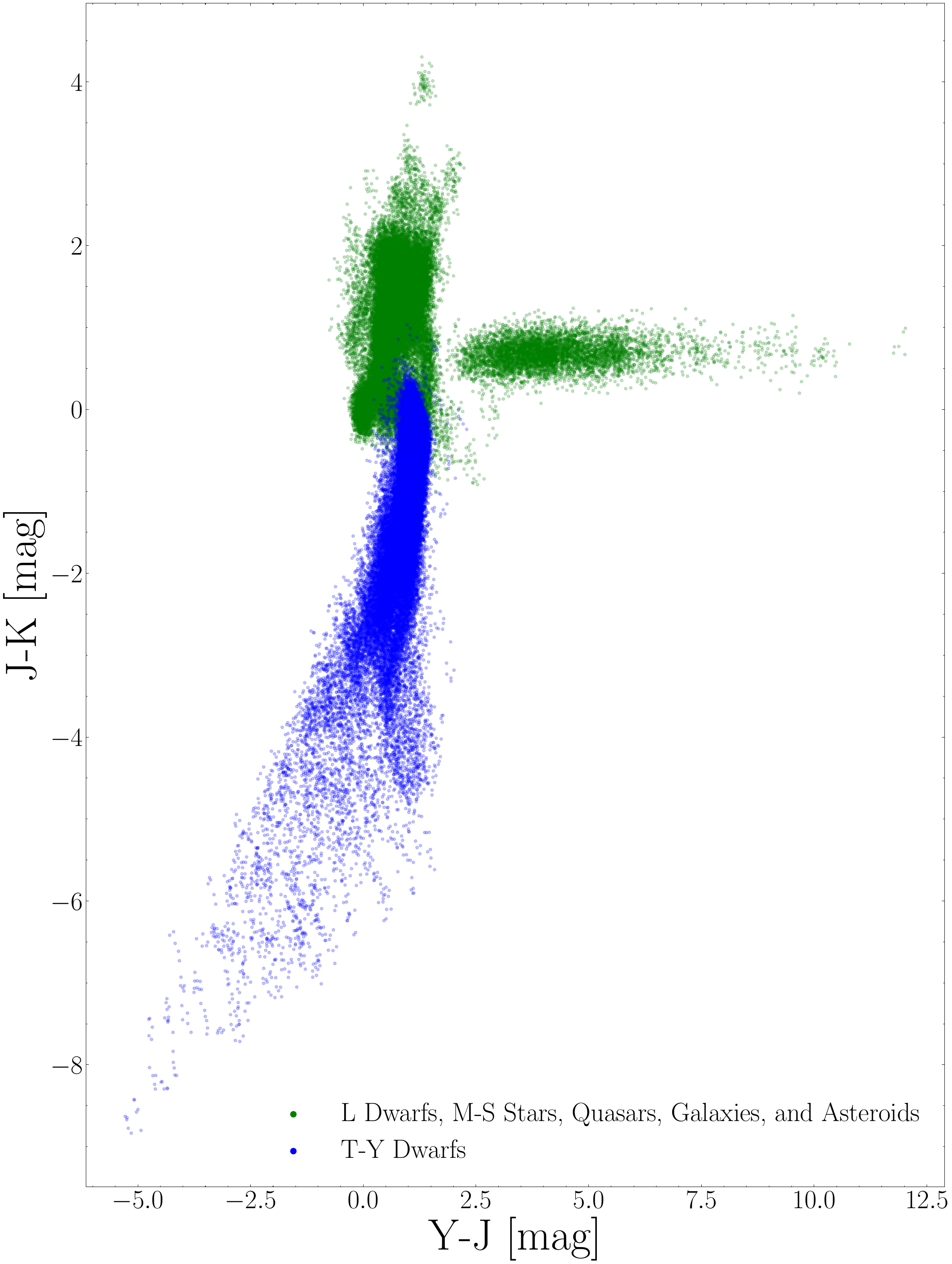}

 \\

\end{tabular}

\caption{Left: JHK color diagram of all synthetic objects. Blue dots are substellar sources and green dots are non-substellar sources. Right: YJK color diagram of all synthetic objects.}
\label{fig:contaminantcolors}.

\end{figure*}

After applying this procedure to all of the contaminant catalogs, they are pooled into one contaminant set. To induce variation in the duplicated objects in step 3 of the class balancing technique and in order to match the length of the total contaminant set to the length of the synthetic TY-dwarf set, the same noise injection technique described in Section ~\ref{sec:dataaug} is applied. The total contaminant dataset is duplicated 11 times to approximately match the length of the synthetic brown dwarf dataset. Color-color diagrams of the datasets are plotted in Figure ~\ref{fig:contaminantcolors}.

\section{The ensemble classifier}
\label{sec:ml}

\begin{figure}
  \resizebox{\hsize}{!}{\includegraphics{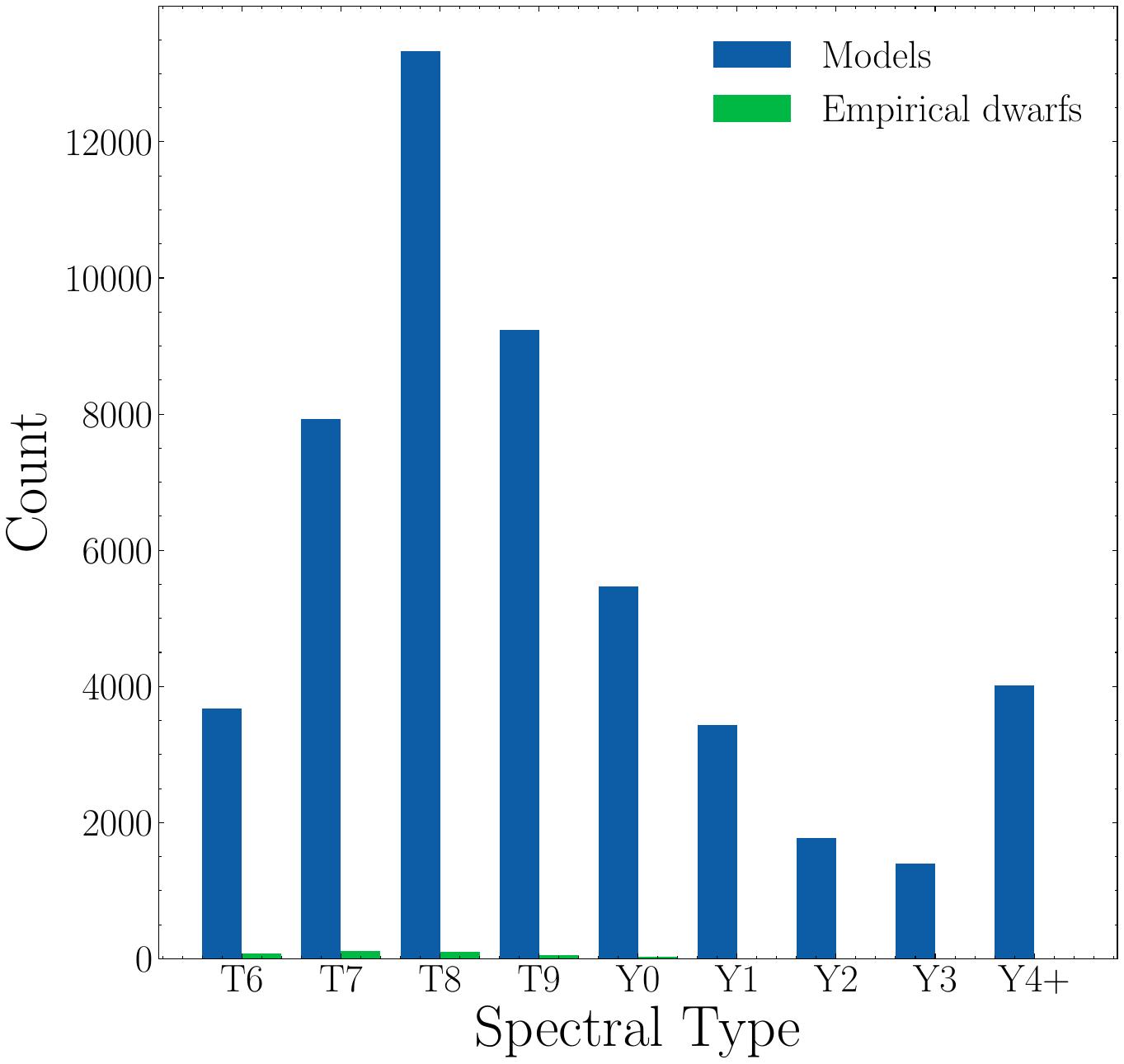}}
  \caption{Histogram of objects for each integer spectral type comparing empirical source counts and the objects in the synthetic data set.}
  \label{fig:datahist}
\end{figure}

The ML-based methods used to classify and detect TY dwarfs are described in this section. Before discussing the classifiers, it is important to acknowledge the difference in training set size between empirical observations and the modified atmospheric models. As shown in Figure ~\ref{fig:datahist}, where the current corpus of UCDs in the SIMBAD is compared to the synthetic dataset in this work, the usage of these models not only offers a significantly larger training dataset, but also offers a slightly more homogeneous coverage of the late brown dwarf spectrum. When comparing the entropy of the two distributions (which both share the same binning and support), the synthetic dataset presents a higher entropy by 0.6 bits, indicating a meaningful shift towards uniformity. It is important to note, though, that particular artifacts at $\ge$Y1 negatively affect this uniformity in the synthetic set (see Section ~\ref{sec:photclass}).

\subsection{A 2-part classifier}
\label{sec:twopart}

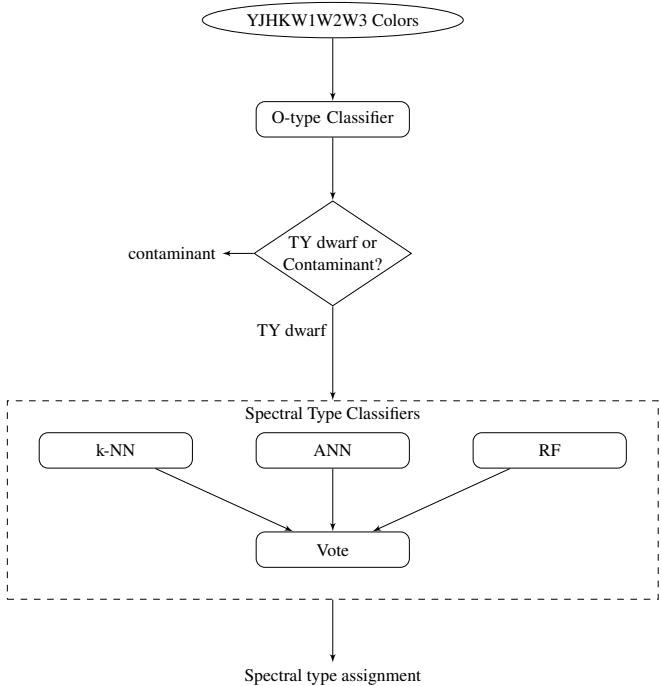
\begin{figure}
\begin{adjustbox}{width=\linewidth,center} 
\begin{tikzpicture}[
    node distance = 1cm,
    auto,
    decision/.style = {diamond, draw, text width=6em, text badly centered, inner sep=0pt, aspect=1.5},
    block/.style = {rectangle, draw, text width=8em, text centered, rounded corners, minimum height=2em},
    line/.style = {draw, -latex'},
    cloud/.style = {draw, ellipse, minimum height=2em},
    group/.style = {draw, rectangle, dashed, inner sep=0.5cm}
]

\node [cloud] (input) {YJHKW1W2W3 Colors};
\node [block, below=of input] (classifier) {O-type Classifier};
\node [decision, below=of classifier, node distance=2.5cm] (decision) {TY dwarf or Contaminant?};

\node [left=0.5cm of decision.west, text width=2cm, align=right] (contaminant) {contaminant};

\begin{scope}[node distance=1cm]
\node [block, below=2cm of decision] (ann) {ANN};
\node [block, left=of ann] (knn) {k-NN};
\node [block, right=of ann] (rf) {RF};

\node [block, below=of ann] (vote) {Vote};
\end{scope}

\node [group, fit=(knn) (rf) (ann) (vote), 
      label={[anchor=north]north:Spectral Type Classifiers}] (stc) {};


\node [below= 1cm of stc] (cut) {Spectral type assignment};

\path [line] (input) -- (classifier);
\path [line] (classifier) -- (decision);
\path [line] (decision.south) -- node [near start, left] {TY dwarf} (stc.north);
\path [line] (knn) -- (vote);
\path [line] (rf) -- (vote);
\path [line] (ann) -- (vote);
\path [line] (stc) -- (cut);
\path [line] (decision.west) -- (contaminant.east);


\end{tikzpicture}
\end{adjustbox}
\caption{Flowchart of ensemble classifier.}
\label{fig:flowplot}
\end{figure}

To minimize sample contamination from background sources, a 2-part classification scheme is adopted, with the former part being allocated solely to contaminant discrimination and the latter part focusing on source classification. The full structure of the ensemble classifier can be seen in Figure ~\ref{fig:flowplot}. 

The classifiers are trained on all possible pairwise combinations among the seven used bands, resulting in 21 unique features. With 21 color-based features derived from only seven bands, there is potential for multicollinearity. Evidence for this can be seen in Figure ~\ref{fig:featureimportance}, where many of the considered colors have near little impact on the random forest's accuracy. Nevertheless, all 21 features are used to avoid the loss of atmospheric properties not captured by the models (See Section ~\ref{sec:intro} and ~\ref{sec:discussion}). The following sections describe the methods\footnote{Models were either built from the \texttt{Sci-Kit Learn} Python module \citep{pedregosa2011scikit} or manually built through the Pytorch framework \citep{paszke2019pytorch}.} used to detect brown dwarfs and determine their spectral type. 

\subsection{Object-type classifier}
\label{sec:otype}

Trained on the contaminant set from Section ~\ref{sec:contaminants} and the full feature set described in Section ~\ref{sec:twopart}, the Object-type (O-type) model is a binary classifier which distinguishes between background objects and TY dwarfs. To handle the high-dimensional nature of the data, an ANN with four hidden layers–depth chosen via grid search–is utilized. 

For both the O-type and spectral type classifiers, the combined contaminant and TY dwarf dataset is randomly split into train/test/validation set with ratios 0.8, 0.1, and 0.1 respectively. The O-type ANN is iteratively trained for 33 epochs with a binary cross-entropy (BCE) loss term before passing a 10-epoch early-stopping patience threshold for loss decrease. Once stopped, the model weights in the iteration with the best test loss are used for the final O-type classifier. The validation accuracy can be seen in Table ~\ref{table:acc}. 

In comparison to the ANN, a k-NN with $k=3$ and the same training set was also tested. While the k-NN O-type classifier performs better on the validation set than the ANN O-type classifier (See Table ~\ref{table:acc}), it is shown in Section ~\ref{sec:tycat} that the latter performs significantly better with empirical data. This tension signifies that the training data may not completely generalize to the empirical UCDs.

\subsection{Spectral type classifiers}
\label{sec:sptype}

Once potential candidates are identified by the O-type classifier, an ensemble of spectral type classifiers predict the spectral type of the candidate object as described in the section below. 

The ensemble spectral type classifier is composed of three independent ML models—a k-NN, a RF, and an ANN. All models are trained with a 10-epoch early-stopping scheme and a mean-squared-error (MSE) loss function. Section ~\ref{sec:tycat} demonstrates that this ensemble classifier functions better than any of the composite models alone. The definitions of these models and their use in the overall spectral type classifier is discussed below. A discussion of the validation accuracy and performance of these models can be found in Section ~\ref{sec:accuracy}.

\subsubsection{k-NN regression}
\label{sec:knn}

As discussed in Section ~\ref{sec:tycat}, the k-NN O-type classifier is not appropriate for contaminant discrimination due to the significant overlap in color space between the two pertinent datasets. Since this problem is not present in the UCD color-spectral type space, the first component model of the ensemble spectral type classifier is a k-NN regressor with $k=3$, which performs a local interpolation of the closest 3 neighboring models to assign an object a spectral type. 

\subsubsection{Random forest regression}
\label{sec:randomforest}

\begin{figure}
  \resizebox{\hsize}{!}{\includegraphics{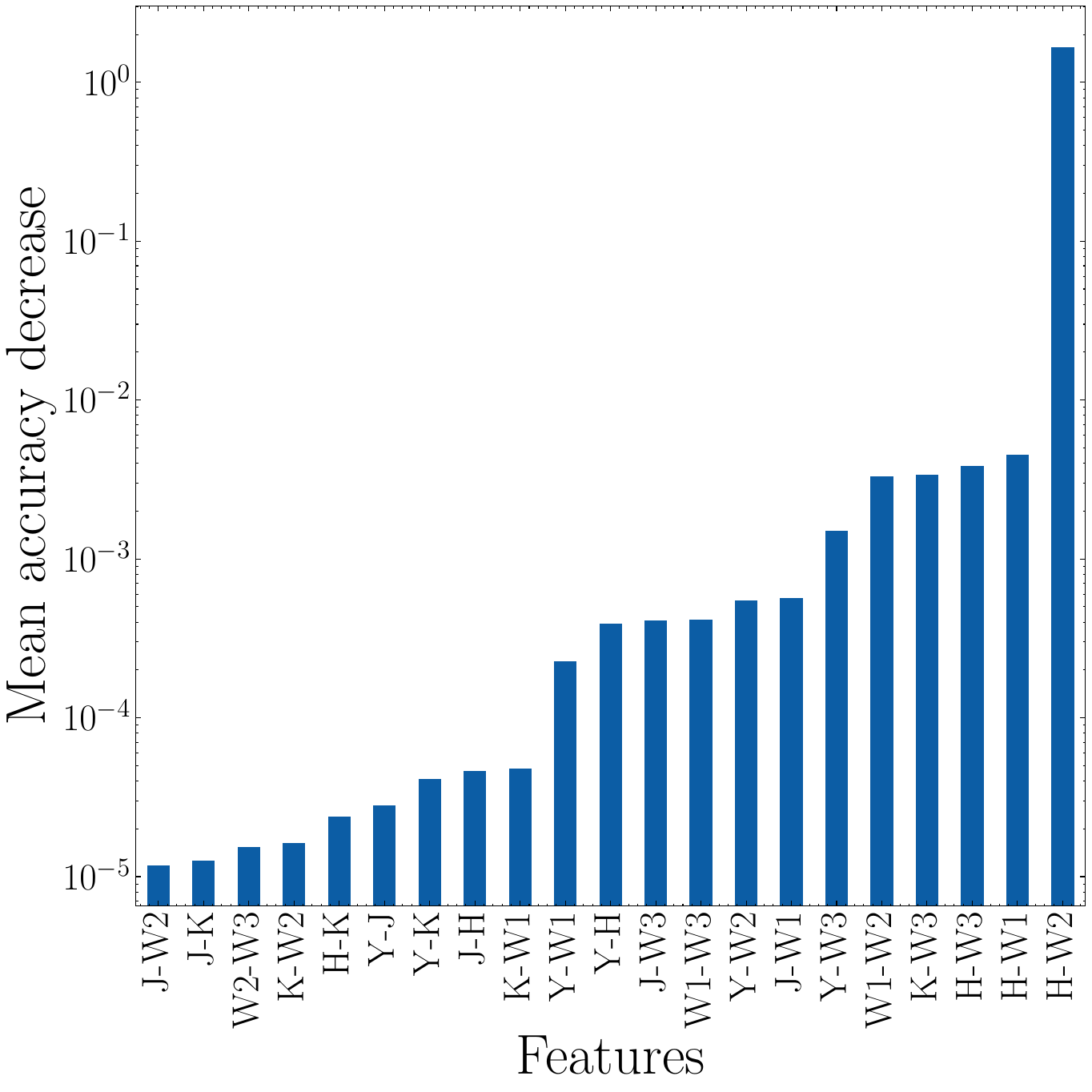}}
  \caption{Relative feature importance of each trained feature in the spectral type random forest classifier. The mean accuracy decreases are derived with permutation feature importance \protect\citep{altmann2010permutation}, where the overall accuracy loss of the model per color is observed by individually randomizing each input feature.}
  \label{fig:featureimportance}
\end{figure}

RFs \citep{ho1995random, breiman2001random} are a type of ensemble classification algorithm, specifically relying on the average of multiple decision trees–in the case of regression–to predict continuous variables. They have been previously utilized in spectral type subgroup assignment in \cite{gong2022applying, gutierrez2022applying}. Rather than utilizing RFs to assign dwarfs to spectral type subgroups, this work uses them to assign discrete spectral types to detected UCDs as a part of the ensemble classifier.  In order to prevent overfitting of the model, a max tree depth of 7 is used. The relative importance of each trained feature is shown in Figure ~\ref{fig:featureimportance}. The high importance given to H-band colors is likely an effect of the use of H as the reference band during photometric template fitting (see Section ~\ref{sec:polynomials}). There is also no apparent agreement with the results of \cite{gong2022applying}, who find W1-W2 and K-W2 as two of the most important features in their spectral type classifier, though it is important to acknowledge that this work does not include two of the bands used in their analysis (see Section ~\ref{sec:valacc} for more discussion on this point) and does not cover their entire spectral type range.

\subsubsection{ANN}
\label{sec:sptypeANN}

In addition to the ANN used in the O-type classifier, an ANN is also utilized as a sub-component of the spectral type classifier. This ANN has the same four-hidden-layer structure as the O-type classifier but is instead trained to predict continuous spectral type outputs. 

\subsection{Ensemble classification}
\label{sec:ensemble}

The aforementioned classifiers are compiled into one ensemble classifier. Below is the pipeline an object would pass through in the classification sequence (also see Fig ~\ref{fig:flowplot}).

\begin{enumerate}
    \item Classify the object type (substellar or non-substellar) of a source with the 21-color feature set and O-type classifier.
    \item If the source is a late-T or Y dwarf, predict its spectral type with the k-NN algorithm, RF regressor, and ANN classifier.  
    \item Take the predicted spectral type as the average of the predictions of the three ML models. 
\end{enumerate}

\section{Results}
\label{sec:results}

The validation accuracy and the performance of the ensemble classifier on various empirical datasets is discussed here. This work's attempts to search for candidate TY dwarfs in the UKIDSS Ultra Deep Survey \citep[UDS;][]{lawrence2007ukirt} and in a region around the Pisces constellation is also covered in this section.

\subsection{Accuracy}
\label{sec:accuracy}

\subsubsection{Validation accuracy}
\label{sec:valacc}

The model scores for the validation set are shown in Table ~\ref{table:acc}. The $r^2$ and RMSE values indicate the correlation and standard deviation of the residuals respectively between the validation set and predicted values. For the O-type classifiers specifically, an F1 score is computed, which is given by 

\begin{equation}
F1 = \frac{2*\mathrm{precision}*\mathrm{recall}}{\mathrm{precision} + \mathrm{recall}}.
\end{equation},

where precision (the positive predictive value) is given by $TP/(TP+FP)$ and recall (the true positive rate) is given by $TP/(TP+FN)$; TP, FP, TN, and FN correspond to true positive, false positive, true negative, and false negative counts respectively. 

\begin{table*}
\caption{Accuracy metrics for each of the ML schemes. For the regression classifiers, accuracies are given in the correlation coefficients of fitted solutions and root mean squared errors (RMSEs). The F1-score is also shown for the binary-class O-type classifiers.}              
\label{table:acc}      
\centering    
\begin{tabular}{c c c c c c c}          
\hline

\hspace{1cm} & k-NN (O-type) & ANN (O-type) & k-NN (sp-type) & RF & ANN (sp-type) & Ensemble \\
\hline
Correlation Coefficient ($r^2$) & 0.994 & 0.973 & 0.995 & 0.994 & 0.989 & 0.998 \\
RMSE & 0.037 & 0.083 & 0.143 & 0.162 & 0.210 & 0.113 \\
F1-score & 0.998 & 0.924 & - & - & -\\
\hline
\end{tabular}
\centering
\end{table*}

Note that the validation scores for the O-type ANN in Table ~\ref{table:acc} are universally worse than those for the O-type k-NN, if only by a minor margin. However, as noted in Section ~\ref{sec:tycat}, the ANN displays both higher recall and higher precision than the k-NN on empirical data, which is why it is used as the only O-type classifier when searching for UCDs in empirical surveys. Nonetheless, taking into consideration only the validation scores, the both the RF and the k-NN O-type F1-scores agree with the $\sim99\%$ validation metrics of the RF classifiers in \cite{gong2022applying, gutierrez2022applying} that discriminate between UCDs and background sources\footnote{Note that any direct comparison between the results of this study and previous studies attempting to use ML for UCD identification/classification is difficult due to a) the different ranges of targeted UCD spectral types, b) the different coarseness of spectral type binning, and c) the different training sets / contaminant modeling.}. While \cite{gong2022applying} finds the \emph{Sloan Digital Sky Survey} \citep[SDSS;][]{york2000sloan} $i-z$ color feature as most important in object-type prediction, this work does not use  \emph{i} and \emph{z} bands yet achieves similar accuracy regimes in accuracy; this is likely due to the optically-faint nature of late T-Y dwarfs in comparison to earlier spectral types, as high-$z$ background sources also appear red or undetected in the optical. 

On the spectral type classification side, these results also compare favorably with those of \cite{gong2022applying, gutierrez2022applying}, who find $\sim97\%$ accuracy when classifying M5-T9.9 dwarfs into spectral type bins $\sim5$ subtypes wide. While a classification metric is meaningless for the spectral type assignment in this work since the predicted values are essentially continuous instead of being sorted into wide spectral type bins, the high validation metrics–especially for the ensemble classifier–indicate a similarly strong correspondence. The spectral type classifier is further validated in the following section with known UCD catalogs.

\subsubsection{Testing in empirical UCD catalogs}
\label{sec:tycat}

To gauge the applicability of these synthetically-trained ML models to empirical data, they are applied to catalogs of known UCDs. The first of these is the full set of TY dwarfs from \cite{leggett2017type}. The ANN O-type classifier is able to recover all 33 TY dwarfs, whereas the k-NN O-type classifier can only recover 32, failing at the coolest dwarf in the dataset which has a spectral type of Y1.5. Overall, the ensemble classifier achieves a mean spectral type classification offset of $0.35\pm0.37$ subtypes and a RMSE of 0.50 from the given spectral types in the catalog. 

To compare this value to the expected difference, I assume a true spectral type value \( x \) uniformly distributed over a bin of width 0.5. I then assume two perfect estimators: \( E_{0.5}(x) \), the ground truth-like estimator which rounds to the nearest subtype multiple of 0.5, and \( E_{0.1}(x) \), the classifier-like estimator which rounds to the nearest subtype multiple of 0.1. To calculate the expected absolute difference $\mathbb{E}[|E_1(x) - E_5(x)|]$, the absolute values of the five possible differences (\( \{-0.2, -0.1, 0.0, 0.1, 0.2\} \)) are averaged, yielding 0.12 as the expected offset. The mean offset being higher than this predicted value may indicate a systemic difference between the synthetic models and empirical data. Similarly, the RMSE is higher than that of any of the spectral type classifiers in the validation set (see Table ~\ref{table:acc}), pointing towards a similar bias. The cause of this offset is discussed in Section ~\ref{sec:discussion}. Nevertheless, these results still compare favorably with the 0.64 and 1.3 average subtype offsets found in \cite{gutierrez2022applying} and \cite{gong2022applying} respectively. The 33 dwarfs and their assigned spectral types are shown in Table ~\ref{table:leggettdwarfs}.

\begin{table*}
\caption{Classification of TY dwarfs from \protect\cite{leggett2017type}. All the dwarfs are recovered by the O-type classifier and the average offset between the ground truth and predicted spectral type is 0.35 subtypes.}  
\label{table:leggettdwarfs}
\begin{tabular}{ccccccc}
\hline
Name & SpT & Jmag & Hmag & Kmag & W1--W2 & Predicted SpT \\
\hline
2MASSI J0243137-245329 & T6.0 & $15.13\pm0.03$ & $15.39\pm0.03$ & $15.34\pm0.03$ & $1.72\pm0.06$ & T6.5 \\
2MASSI J0937347+293142 & T6.0 & $14.29\pm0.03$ & $14.67\pm0.05$ & $15.39\pm0.06$ & $2.42\pm0.05$ & T7.5 \\
ULAS J115229.67+035927.2 & T6.0 & $17.28\pm0.02$ & $17.70\pm0.05$ & $17.77\pm0.12$ & $1.71\pm0.19$ & T6.7 \\
2MASS J12255432-2739466 & T6.0 & $14.88\pm0.03$ & $15.17\pm0.03$ & $15.28\pm0.03$ & $1.93\pm0.06$ & T6.4 \\
SDSSp J162414.37+002915.6 & T6.0 & $15.20\pm0.05$ & $15.48\pm0.05$ & $15.61\pm0.05$ & $2.07\pm0.07$ & T6.3 \\
2MASS J12373919+6526148 & T6.5 & $15.6\pm0.1$ & $15.9\pm0.1$ & $16.4\pm0.1$ & $2.45\pm0.07$ & T7.5 \\
SDSSp J134646.45-003150.4 & T6.5 & $15.64\pm0.01$ & $15.97\pm0.01$ & $15.96\pm0.02$ & $1.77\pm0.07$ & T6.3 \\
SDSS J175805.46+463311.9 & T6.5 & $15.86\pm0.03$ & $16.20\pm0.03$ & $16.12\pm0.03$ & $1.85\pm0.06$ & T6.5 \\
2MASS J00501994-3322402 & T7.0 & $15.6\pm0.1$ & $16.0\pm0.1$ & $15.9\pm0.1$ & $2.02\pm0.07$ & T6.8 \\
2MASSI J0727182+171001 & T7.0 & $15.19\pm0.03$ & $15.67\pm0.03$ & $15.69\pm0.03$ & $2.23\pm0.07$ & T6.9 \\
ULAS J144902.02+114711.4 & T7.0 & $17.36\pm0.02$ & $17.73\pm0.07$ & $18.10\pm0.15$ & $2.21\pm0.16$ & T7.1 \\
2MASS J11145133-2618235 & T7.5 & $15.52\pm0.05$ & $15.82\pm0.05$ & $16.54\pm0.05$ & $2.98\pm0.06$ & T8.0 \\
2MASSI J1217110-031113 & T7.5 & $15.56\pm0.03$ & $15.98\pm0.03$ & $15.92\pm0.03$ & $2.06\pm0.07$ & T7.1 \\
ULAS J141623.94+134836.3 & T7.5 & $17.34\pm0.02$ & $17.65\pm0.03$ & $18.93\pm0.02$ & $3.21\pm0.23$ & T8.9 \\
2MASS J14571496-2121477 & T7.5 & $14.82\pm0.05$ & $15.28\pm0.05$ & $15.52\pm0.05$ & $2.8\pm0.06$ & T7.6 \\
WISEP J025409.45+022359.1 & T8.0 & $15.92\pm0.02$ & $16.29\pm0.02$ & $16.73\pm0.05$ & $3.05\pm0.08$ & T7.8 \\
2MASSI J0415195-093506 & T8.0 & $15.32\pm0.03$ & $15.70\pm0.03$ & $15.83\pm0.03$ & $2.85\pm0.07$ & T7.8 \\
2MASS J07290002-3954043 & T8.0 & $15.6\pm0.1$ & $16.0\pm0.1$ & $16.6\pm0.1$ & $2.32\pm0.05$ & T7.5 \\
2MASS J09393548-2448279 & T8.0 & $15.61\pm0.09$ & $15.94\pm0.09$ & $16.83\pm0.09$ & $3.27\pm0.05$ & T8.6 \\
ULAS J130041.74+122114.7 & T8.0 & $16.69\pm0.01$ & $17.01\pm0.04$ & $16.90\pm0.06$ & $2.19\pm0.10$ & T7.4 \\
ULAS J130217.21+130851.2 & T8.0 & $18.11\pm0.05$ & $18.60\pm0.05$ & $18.28\pm0.03$ & $2.82\pm0.3$ & T8.0 \\
CFBDS J005910.90-011401.3 & T8.5 & $18.06\pm0.05$ & $18.27\pm0.05$ & $18.71\pm0.05$ & $3.17\pm0.16$ & T8.7 \\
ULAS J133553.45+113005.2 & T8.5 & $17.90\pm0.01$ & $18.25\pm0.01$ & $18.28\pm0.03$ & $3.05\pm0.14$ & T8.6 \\
WISE J000517.48+373720.5 & T9.0 & $17.59\pm0.02$ & $17.98\pm0.03$ & $17.99\pm0.03$ & $3.47\pm0.12$ & T8.7 \\
UGPS J072227.51-054031.2 & T9.0 & $16.52\pm0.02$ & $16.90\pm0.02$ & $17.07\pm0.08$ & $3.05\pm0.07$ & T9.0 \\
WISEP J075108.79-763449.6 & T9.0 & $19.4\pm0.1$ & $19.7\pm0.1$ & $20.0\pm0.2$ & $2.42\pm0.11$ & T9.0 \\
WISEP J121756.91+162640.2 & T9.0 & $17.83\pm0.02$ & $18.18\pm0.05$ & $18.80\pm0.04$ & $3.42\pm0.11$ & T9.2 \\
WISE J081117.81-805141.3 & T9.5 & $19.65\pm0.07$ & $19.99\pm0.14$ & $20.49\pm0.2$ & $2.43\pm0.12$ & T9.6 \\
WISE J035934.06-540154.6 & Y0.0 & $21.5\pm0.1$ & $21.7\pm0.2$ & $22.8\pm0.3$ & $3.7\pm0.3$ & Y0.1 \\
WISEP J173835.53+273258.9 & Y0.0 & $19.58\pm0.04$ & $20.24\pm0.08$ & $20.58\pm0.1$ & $3.2\pm0.2$ & T9.7 \\
WISEP J205628.90+145953.3 & Y0.0 & $19.43\pm0.04$ & $19.96\pm0.04$ & $20.01\pm0.06$ & $2.64\pm0.12$ & Y0.0 \\
WISEP J140518.40+553421.4 & Y0.5 & $21.06\pm0.06$ & $21.41\pm0.08$ & $21.61\pm0.12$ & $4.7\pm0.4$ & Y0.8 \\
WISEP J182831.08+265037.8 & Y1.5 & $23.5\pm0.2$ & $22.7\pm0.1$ & $23.5\pm0.4$ & $3.0\pm0.3$ & Y1.4 \\
\hline
\end{tabular}
\end{table*}

The classifier is also validated against the set of M,L, and T dwarfs from \cite{best2018photometry} cross-matched with UKIDSS LAS DR9 with a matching radius of 3.5 arcsec. The classifier is run on the cross-matched catalog of 4983 objects, recovering all late-T dwarfs and 4 of the 4953 M/L/early-T contaminants (0.08\%). The classification of all late T dwarfs in the catalog is shown in Table ~\ref{table:bestdwarfs}. The data does not support any particular bias against peculiar dwarfs as reported in \cite{gutierrez2022applying}. The k-NN O-type classifier, meanwhile, also recovers the 8 UCDs from the \cite{best2018photometry} set but also misclassifies 38 M/L dwarfs as TY dwarfs, indicating the higher precision of the ANN O-type classifier on empirical data. 

\begin{table*}
\caption{Classification of T dwarfs from \protect\cite{best2018photometry}. All T dwarfs are recovered by the O-type classifier.}  
\label{table:bestdwarfs}
\begin{tabular}{ccccccc}
\hline
Name & SpT & Jmag & Hmag & Kmag & W1--W2 & Predicted SpT \\
\hline
2MASS J00345157+0523050 & T6.5 & $15.54\pm0.01$ & $15.44\pm0.01$ & $16.07\pm0.03$ & $2.54\pm0.07$ & T7.2 \\
SDSSp J134646.45-003150.4 & T6.5 & $16.00\pm0.01$ & $15.46\pm0.01$ & $15.96\pm0.02$ & $1.77\pm0.07$ & T6.0 \\
WISE J230133.32+021635.0 & T6.5 & $16.71\pm0.01$ & $16.09\pm0.03$ & $16.87\pm0.05$ & $1.86\pm0.1$ & T6.8 \\
ULAS J234228.97+085620.1 & T6.5 & $16.73\pm0.01$ & $16.26\pm0.03$ & $16.98\pm0.07$ & $2.02\pm0.1$ & T6.4 \\
WISE J004024.88+090054.8 & T7.0 & $16.50\pm0.01$ & $16.54\pm0.02$ & $16.55\pm0.05$ & $2.16\pm0.1$ & T7.0 \\
SDSS J150411.63+102718.4 & T7.0 & $17.03\pm0.01$ & $16.91\pm0.05$ & $17.12\pm0.08$ & $2.16\pm0.1$ & T7.2 \\
ULAS J141623.94+134836.3 & T7.5 & $17.63\pm0.02$ & $17.55\pm0.03$ & $18.93\pm0.2$ & $3.21\pm0.2$ & T8.7 \\
WISEPC J222623.05+044003.9 & T8.0 & $17.02\pm0.02$ & $17.30\pm0.07$ & $17.24\pm0.09$ & $2.35\pm0.2$ & T7.1 \\
\hline
\end{tabular}
\end{table*}

\begin{figure}
  \resizebox{\hsize}{!}{\includegraphics{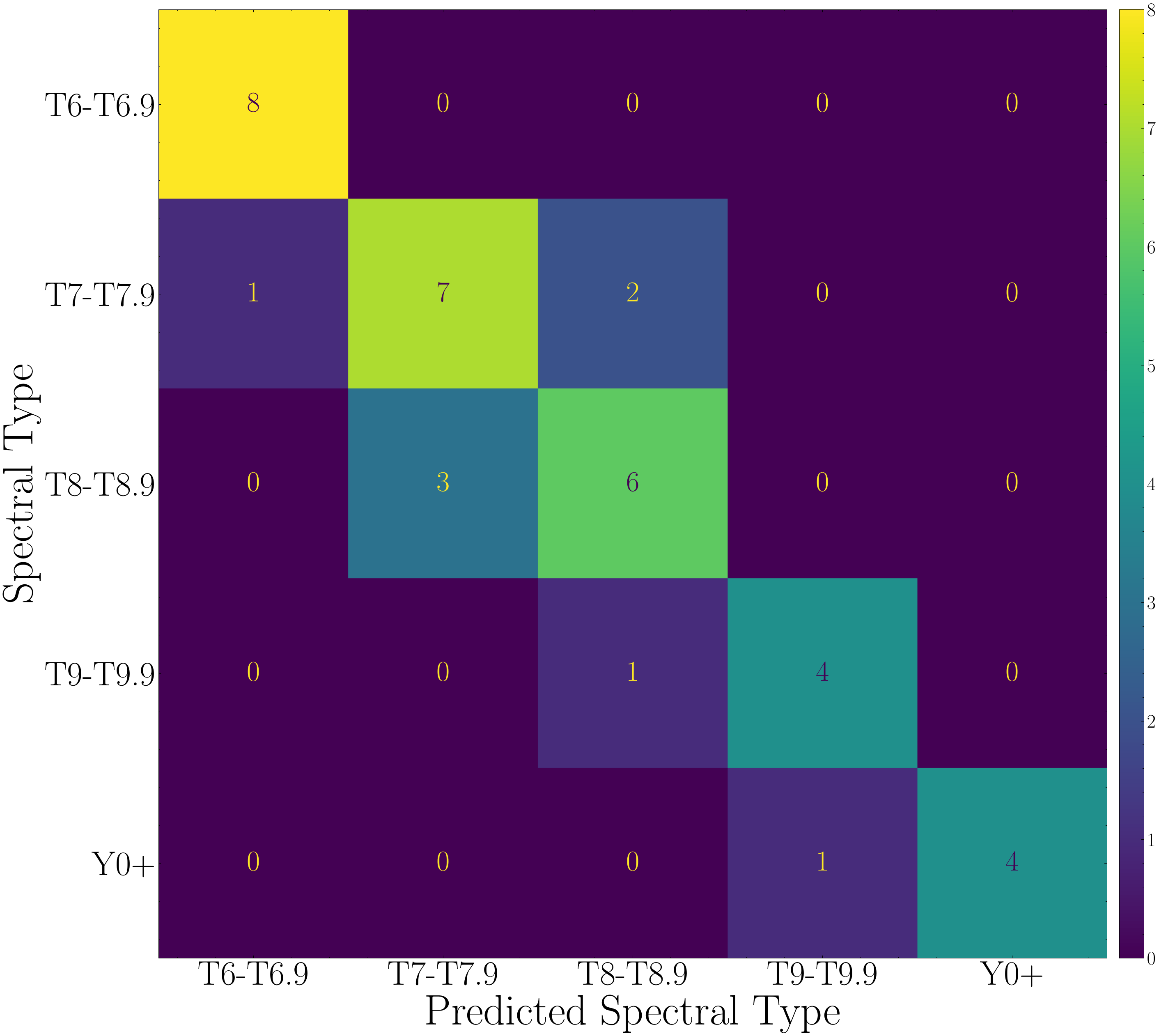}}
  \caption{A binned confusion matrix of the original spectral types and the predicted spectral types from the combined $\ge$T6 dwarfs of Table ~\ref{table:leggettdwarfs} and Table ~\ref{table:bestdwarfs}.}
  \label{fig:sptypeconf}
\end{figure}

The binned spectral types and the classified spectral types are shown in the form of a confusion matrix in Figure ~\ref{fig:sptypeconf}. As indicated by the matrix, there is relatively consistent agreement between the spectral type and the predicted spectral type. 

\subsection{Search for TY dwarfs in UDS}
\label{sec:uds}

To evaluate the performance of the ensemble classifier in a source discovery context, the UKIDSS UDS field \citep{lawrence2007ukirt} is selected. I start with a cross-matched 12-band photometric UDS catalog from Almaini et al., Hartley et al. (in preparation) that contains UKIRT JHK data with cross-matched VIRCAM Y-band photometry. The catalog coverage extends over the entire 0.8 sq. degree UDS field with $95\%$ completeness at K = 25.0 (AB). Any sources below the UDS $5\sigma$ AB magnitude limits, $m_Y=24.6,m_J=25.6,m_H=25.1,m_K=25.3$, are cut.

Further, sources with a \texttt{CLASS\_STAR} $\ge 0.8$, marked as \texttt{SATURATED}, or with \texttt{GOOD GALAXIES} or \texttt{BEST GALAXIES} tags, are removed from the catalog. Finally, after cross-matching with AllWISE with a 3.5 arcsec radius, the catalog is further pruned with a 11 mag saturation limit in all bands. Point sources are isolated by selecting sources with the AllWISE \texttt{ex} tag equal to 0. This results in a final searchable catalog of 299 sources. 

After applying the O-type classifier, no UCDs are found and all sources are classified as contaminants. This appears to contradict theoretical halo surface densities at this depth, as the most conservative model from \cite{aganze2022beyond}, who model expected counts from a JWST survey at similar depths to UDS ($m_\text{F140W}=25.0$), finds 0.00283 arcmin$^{-2}$ for the density of T dwarfs ($\sim6$ expected in this field)–it is noted that no early-T dwarfs have been detected in the field either to augment these counts. Similarly, these results are in tension with the predictions of \cite{deacon2006possiblity}, who expect 45-208 late-T dwarfs and 3-58 Y dwarfs within a shallower ($m_K=23$) survey in the UDS field. Nevertheless, the non-detection corroborates previous searches, which have also failed to identify UCDs in the UDS field \citep{lodieu2009two, skrzypek2014brown, aganze2022beyond}. Taken together, these results may suggest a genuine paucity of UCDs in this region, potentially due to variations in Galactic structure, such as a local underdensity or a lower scale height of the substellar population than previously assumed. Alternatively, they may point to limitations in current models of UCD spatial distribution.

\subsubsection{TY dwarfs in the constellation of Pisces}
\label{sec:kapgem}

This work also searches for UCDs in a $1.5^{\circ}$ radius region in the constellation of Pisces centered at 00 08 11.456 +01 20 2.07. This region, when cross-matched with UKIDSS LAS DR11, AllWISE, and the SIMBAD catalog (all with a radius of 3.5 arcsec), contains 612 cataloged objects. Sources are selected with $5\sigma$ limiting magnitudes of $m_Y=20.2,m_J=20.0,m_H=18.8,m_K=18.2$.

There is 1 confirmed T6.5 dwarf in this region. The O-type classifier is able to recover one T dwarf and one additional object not cataloged with SIMBAD. This additional object is classified by the ensemble model as a T8.2 dwarf and is shown in Table ~\ref{table:kapgem}. Extrapolating from the results of \cite{deacon2006possiblity}, 1-4 late-T dwarfs are expected to be found in this region at LAS depths, which aligns with these results (2 dwarfs–1 catalogued, 1 uncatalogued). The results agreeing with surface density predictions in this case may indicate an unique feature in the UDS region (see Section ~\ref{sec:uds}). 

\begin{table*}
\caption{Identified T dwarfs in a $1.5^{\circ}$ region in the constellation of Pisces.}
\label{table:kapgem}
\centering
\begin{tabular}{cccccccc}
\hline
Name & SIMBAD O-type & $J$ & $H$ & $K$ & $W1-W2$ & SIMBAD SpT & Predicted SpT \\
\hline
ULAS J000844.34+012729.4 & BrownD* & $16.99 \pm 0.02$ & $17.40 \pm 0.06$ & $17.54 \pm 0.10$ & $2.19 \pm 0.32$ & T6.5 & T7.4 \\
AllWISE J000711.42+013707.9 & - & $18.74 \pm 0.10$ & $18.39 \pm 0.13$ & $18.12 \pm 0.16$ & $2.70 \pm 0.59$ & - & T8.2 \\
\hline
\end{tabular}
\end{table*}

\subsubsection{Statistical fitting}
\label{sec:statfit}

\begin{figure*}
  \resizebox{\hsize}{!}{\includegraphics{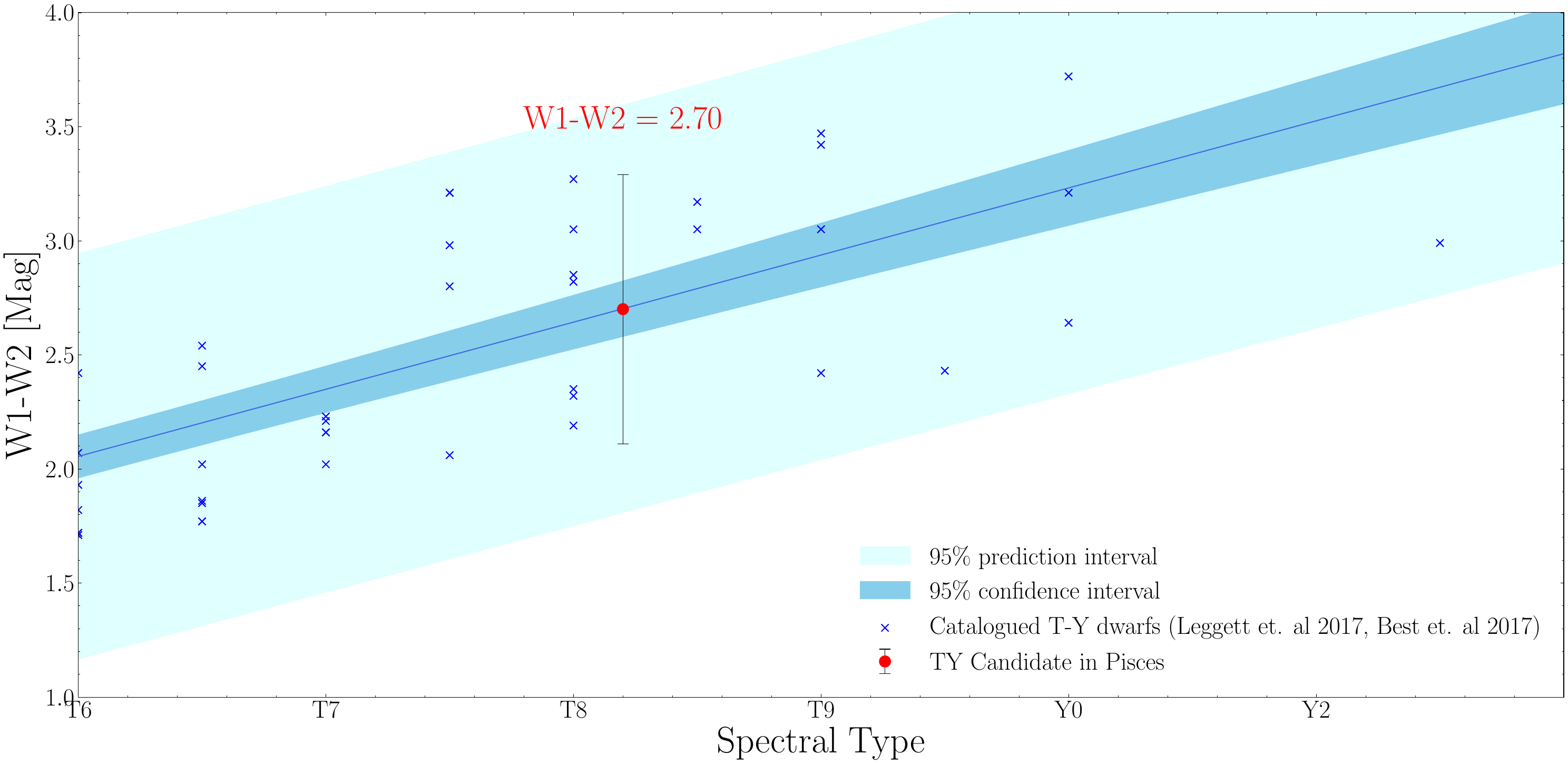}}
  \caption{Spectral type-color graph of the discovered candidate in Pisces. The confidence and prediction intervals are determined from the blue crosses, which are empirical brown dwarf colors sourced from \protect\cite{leggett2017type}. Errorbars for the empirical dwarf colors are omitted for clarity.}
  \label{fig:statfit}
\end{figure*}

To further verify the uncataloged positive candidate detected in the constellation of Pisces, its W1-W2 color is compared against a color prediction interval derived from the dwarfs of \cite{leggett2017type}. The prediction interval is given by:

\begin{equation}
    X = \hat{y} \pm t \cdot s_e \sqrt{1 + \frac{1}{n} + \frac{(x - \bar{x})^2}{\sum_{i=1}^n (x_i - \bar{x})^2}}
\end{equation}

where $\hat{y}$ is the predicted W1-W2 color, the test statistic $t=2$, the standard error of the W1-W2 color $s_e=0.389$, and the number of dwarfs in the \cite{leggett2017type} set $n=70$. The prediction interval and the T8.2 candidate are shown in Figure ~\ref{fig:statfit}. Since the W1-W2 color of the Pisces candidate falls within this interval, it is kept. 

\section{Discussion}
\label{sec:discussion}

\subsection{Caveats}

Overall, this technique hinges on the assumption that the atmosphere models are strongly indicative of properties present in real life. For one, there is doubt as to if the models can predict the brightening of H-band magnitudes in late TY dwarfs \citep{phillips2020new}. As discussed in Section ~\ref{sec:intro}, models like ATMO 2020 and Sonora Bobcat are unable to accurately replicate the colors of late T-dwarfs and early Y-dwarfs. \cite{zhang2021uniform} identifies a $2-4\%$ systematic discrepancy between Sonora-Bobcat spectra and the colors of empirical dwarfs. However, a scatter larger than this discrepancy is applied during the training set augmentation in Section ~\ref{sec:dataaug}, so this likely does not play a large role in the observed errors. Nevertheless, in order to more effectively use atmospheric models to train for the colors of late UCDs, further progress is needed to emulate the effects of processes not addressed by standard radiative-convective models \citep{phillips2020new}. 

Evidence for such a systemic offset between the training sets and empirical UCDs is found in the ensemble spectral type predictions for the \cite{leggett2017type} dwarfs in Section ~\ref{sec:tycat} (See Table ~\ref{table:leggettdwarfs}). The average absolute offset between the ensemble spectral type predictions and the given subtypes is higher than the expected difference if any such offset was caused solely by the binning resolutions of the two spectral type sets. This likely indicates a systematic difference between the model set and empirical dwarfs. Figure ~\ref{fig:sptypeconf} indicates this difference predominently arises from the missclassification of T6-T8.9 dwarfs, as the spectral types of later dwarfs are in good agreement with the types assigned by \cite{leggett2017type}. This work hypothesizes that this effect is due to the large number of atmospheric models concentrated at the T6-T8.9 subtypes, which consist $58\%$ of the total synthetic dwarf set. Such a dense color-color space in concordance with the noise augmentation technique discussed in Section ~\ref{sec:dataaug} is the likely culprit, leading to significant subtype confusion at the early end of the considered spectral type range. In the future, a class balancing approach would likely solve this issue. 

Nevertheless, the recovery of one new UCD candidate at T8.2 that is in good agreement with W1-W2 colors expected from known empirical dwarf catalogs (See Figure ~\ref{fig:statfit}) and the rejection of most contaminants within empirical catalogs indicates that a model-inspired approach to finding UCDs holds promise.

\section{Conclusion}
\label{sec:conclusion}

This work presents a novel atmospheric-model-based classification system that uses an ensemble of machine learning models to search for late-T and Y dwarfs. This system was trained on models from the \emph{ATMO 2020} and \emph{Sonora Bobcat} sets. Using these models to create an artificial catalog of UCDs, they were photometrically assigned near-IR spectral types based off of a chi-squared framework from \cite{skrzypek2015photometric}. These synthetic models were subsequently used to train an ensemble of machine learning classifiers, each of which displayed a high degree of accuracy on both synthetic and empirical validation sets. After running this classifier in a region in the constellation of Pisces, one new candidate T8.2 UCD was discovered. Though it is acknowledged that there are flaws with current models in replicating UCD colors, this methodology serves as a proof of concept for the use of atmospheric models for training in the detection of TY dwarfs. 

\section*{Acknowledgements}
I thank Dr. Patrick Treuthardt, Dr. Michael Cushing, and the anonymous reviewer for their invaluable suggestions and feedback. This publication makes use of data products from the Wide-field Infrared Survey Explorer, which is a joint project of the University of California, Los Angeles, and the Jet Propulsion Laboratory/California Institute of Technology, funded by the National Aeronautics and Space Administration. This research has also made use of the SIMBAD database, Vizier catalogue, and the cross-match service operated at CDS, Strasbourg, France. The UKIDSS project is defined in \cite{lawrence2007ukirt}. UKIDSS uses the UKIRT Wide Field Camera (WFCAM; \cite{casali2007ukirt}) and a photometric system described in \cite{hewett2006ukirt}. The pipeline processing and science archive are described in Irwin et al. (in preparation) and \cite{hambly2008wfcam}. I have used data from the 11th data release. Additionally, this paper uses data from the VISTA Hemisphere Survey ESO programme ID: 179.A-2010 (PI. McMahon).

\section*{Data Availability}
\label{sec:dataavail}

The ATMO2020 models \citep{phillips2020new} used to simulate TY-dwarf atmospheres for contaminant modeling are available at \url{https://perso.ens-lyon.fr/isabelle.baraffe/ATMO2020/}. The Sonora Bobcat models \citep{marley2021sonora} are available for download at \url{https://zenodo.org/records/5063476}. The code for the ensemble classifier created here is available upon request to the author.

\bibliographystyle{mnras}
\bibliography{bibliography-copy}

\bsp	
\label{lastpage}
\end{document}